\documentclass[a4paper, 11pt]{article}
\usepackage[utf8]{inputenc}
\usepackage{fullpage}
\usepackage[margin=1in, letterpaper]{geometry}
\usepackage{my_package}
\usepackage{multirow}
\usepackage{microtype}
\usepackage{XCharter}
\usepackage{dirtytalk}
\usepackage{thm-restate}
\usepackage{enumitem}
\usepackage{fancyhdr}
\usepackage[normalem]{ulem}                 

\newcommand{\SLS}{\textsc{Sliding Local-Search}}
\newcommand{\SLSS}{\textsc{SlidingLS}}
\newcommand{\ILS}{\textsc{Interval Local-Search}}

\date{}

\title{Better Approximation for Weighted $k$-Matroid Intersection}

\author{
Neta Singer \thanks{Ecole Polytechnique Fédérale de Lausanne
(\href{mailto:neta.singer@epfl.ch}{neta.singer@epfl.ch},
\href{mailto:theophile.thiery@epfl.ch}{theophile.thiery@epfl.ch})}
\and Theophile Thiery \footnotemark[1] \thanks{Supported by the Swiss State Secretariat for Education, Research and Innovation (SERI) under contract number MB22.00054}}

\begin{document}
\maketitle
\pagestyle{plain}

    \begin{abstract}
    We consider the problem of finding an independent set of maximum weight simultaneously contained in $k$ matroids over a common ground set. This \emph{$k$-matroid intersection problem} appears naturally in many contexts, for example in generalizing graph and hypergraph matching problems.
    In this paper, we provide a $(k+1)/(2 \ln 2)$-approximation algorithm for the weighted $k$-matroid intersection problem. This is the first improvement over the longstanding $(k-1)$-guarantee of Lee, Sviridenko and Vondr\'ak (2009). Along the way, we also give the first improvement over \textsc{greedy} for the more general weighted matroid $k$-parity problem.

    Our key innovation lies in a randomized reduction in which we solve almost unweighted instances iteratively.
    This perspective allows us to use insights from the unweighted problem for which Lee, Sviridenko, and Vondr\'ak have designed a $k/2$-approximation algorithm. We analyze this procedure by constructing refined matroid exchanges and leveraging randomness to avoid bad local minima.

    \end{abstract}

    \pagenumbering{roman}
    \thispagestyle{empty}
    \newpage
    \setcounter{tocdepth}{2}{\sffamily\tableofcontents}
    \newpage
    \pagenumbering{arabic}
    \setcounter{page}{1}

	\section{Introduction}
        \label{sec:Introduction}

        $k$-Matroid Intersection is a fundamental question in combinatorial optimization where the goal is to find a maximum weight independent set in the intersection of $k$ matroids defined on a common ground set.
        This problem generalizes many classical optimization tasks, including finding maximum matchings in bipartite graphs and $k$-partite hypergraphs. Beyond its versatile formulation, it provides a rich mathematical environment capturing the success of popular approximation algorithms heuristics. For $k = 1$, the problem is solvable in polynomial time using \textsc{Greedy}. Moreover, any independence system (or downward closed family of sets) for which \textsc{Greedy} is optimal is a matroid \cite{Rado:1957:Note, Gale:1968:Optimal}. For $k = 2$, Edmonds' celebrated polynomial-time algorithm for matroid intersection, generalizing the Hungarian Method, finds the largest independent set in the intersection of two matroids \cite{Edmonds:2003:Submodular}. \\

        As soon as the intersection is taken over $3$ or more matroids, no algorithm is known to have optimal approximation guarantee.
        For $k \geq 3$, $k$-Matroid Intersection was recently shown to be APX-hard to approximate within a factor $k/12$ unless $\mathbf{NP} \subseteq \mathbf{BPP}$ \cite{Lee:2024:Asymptotically}, improving on the bound of \cite{Hazan-Safra-Schwartz:2006:Complexity}.

        Currently, the best approximation ratio for the \emph{unweighted} problem stands at $k/2$ \cite{Lee:2009:Matroid}. Notably, this result does not extend to settings with arbitrary weights.
        In fact, for weighted $k$-Matroid Intersection Lee, Sviridenko and Vondr\'ak obtained a tight $k-1$-approximation algorithm, which remains the best known to date  \cite{Lee:2010:Submodular}. \\

        Since Lee, Sviridenko and Vondr\'ak's result, most algorithmic research has been devoted to the study of special cases of weighted $k$-Matroid Intersection. A notable case is $k$-Set Packing (technically $k$-Dimensional Matching), which consists of finding a maximum weight matching in a $k$-partite hypergraph.
        After a series of work \cite{Arkin:1998:Local, Chandra:1999:Greedy, Berman:2000:d/2, Berman:2003:Optimizing, Neuwohner:2021:Improved, Neuwohner:2022:Limits, Thiery:2023:Improved, Neuwohner:2023:Passing}, the state-of-the-art approximation algorithm for the weighted problem is equal to $\frac{k}{2.006}$ \cite{Neuwohner:2023:Passing}.
        Despite great algorithmic progress, known algorithms for $k$-Set Packing don’t appear to extend naturally to the more general $k$-Matroid Intersection. The simplest method accomplishing a $(k+1)/2$ approximation factor is due to Berman \cite{Berman:2000:d/2}. This method performs a local-search procedure on an instance with squared-weights and is used and refined in \cite{Berman:2003:Optimizing, Neuwohner:2021:Improved, Neuwohner:2022:Limits, Thiery:2023:Improved, Neuwohner:2023:Passing}. However, their analyses rely on properties that are only true for the intersection of partition matroids.
        This situation highlights our limited understanding of $k$-Matroid Intersection beyond $k$-Set Packing and the need for a clean and universal algorithm for the more general problem. Our main result is a novel algorithm for weighted $k$-Matroid Intersection, which is the first to improve upon the $k-1$ approximation algorithm of \cite{Lee:2010:Submodular}.
        \begin{mythm}
            \label{thm:intro}
            There is $\frac{k+1}{2\ln(2)}$-approximation algorithm for weighted $k$-Matroid Intersection.
        \end{mythm}
        Our algorithm is the first approximation algorithm asymptotically improving over \textsc{Greedy} for weighted $k$-Matroid Intersection.
        This improvement over \textsc{Greedy} is substantial since the ratio $\frac{1}{2\ln(2)} \simeq 0.722$ improves the leading constant to more than halfway between \textsc{Greedy} and the state-of-the-art ratio of $k/2$ for the unweighted setting.
        Our algorithm applies the standard local-search algorithm in an iterative fashion and thus also provides an alternative method to the $w^2$-algorithm of Berman \cite{Berman:2000:d/2} for weighted $k$-Set Packing. We, in fact, design a $\frac{k+1}{2\ln(2)}$-approximation algorithm for a more general problem: weighted Matroid $k$-Parity.
        In this problem, we are given a collection of weighted hyperedges each incident to $k$ vertices, and we aim to find the largest weight hypergraph matching whose groundset is independent in a matroid over the vertices of the hypergraph. $k$-Matroid Intersection is a special case of this problem. Matroid ($2$-)Parity was proposed by Lawler \cite{Lawler:2001:Combinatorial} to generalize non-bipartite matching and $2$-Matroid Intersection.
        It is solvable in polynomial time for \emph{linear matroids} as shown by Lov\'asz in the unweighted setting \cite{Lovasz:1980:Matroid-Matching} and by Iwata and Kobayashi with general weights \cite{Iwata-Kobayashi:2017:Weighted}.
        With an arbitrary matroid, the problem requires exponentially many calls to the independence oracle \cite{Jensen-Korte:1982:Complexity}. This problem has a wide variety of applications across different areas . A non-exhaustive list of applications of this problem include Path Packing, Minimum Pinning Set, Maximum Genus Embedding, and Maximum Planar Subgraph which have connection to approximation algorithms, combinatorial rigidity and electric ciruits. We refer the reader to \cite{Cheung-Lau-Leung:2014:Algebraic} for an in-depth literature review of the problem.
        For \emph{unweighted} Matroid $k$-Parity with $k \geq 3$, Lee, Vondr\'ak and Sviridenko \cite{Lee:2009:Matroid} designed a $\frac{k}{2}$ approximation algorithm. Meanwhile, \textsc{Greedy} is still the best known algorithm for handling the problem with general hyperedge weights, giving a $k$-approximation factor. Our contribution is thus to also give the first improvement over \textsc{Greedy} for weighted Matroid $k$-Parity. This answers the question by Lee, Vondr\'ak and Sviridenko \cite{Lee:2009:Matroid} who asked about the existence of improved local search algorithms for the weighted version of Matroid $k$-Parity.

        \subsection{Our local search algorithm}
        The vast majority of approximation algorithms for $k$-Matroid Intersection and $k$-Set Packing, including ours, use a local-search heuristic as a subroutine. Local-search procedures repeatedly improve a solution $A$ by adding some small number of elements $S \subseteq V \bb A$ and removing \emph{conflicting} elements $T \subseteq A$, while preserving the independence of the solution. The pair $(S, T)$ is called a \emph{swap}.
        The algorithm terminates when there is no \emph{improving swap}. The absence of improving swaps then allows to compare the value of the algorithm's output $A$ with the value of an optimal solution. The approximation ratio is dictated by the size of the local search swaps. For example, if $A$ is obtained using \textsc{Greedy} then there is no improving swap of size $1$, and implies that \textsc{Greedy} is a $k$-approximation algorithm. Prior improvements over the factor $k$ were obtained by looking at larger swaps as in \cite{Lee:2009:Matroid, Lee:2010:Submodular} for $k$-Matroid Intersection and \cite{Hurkens-Schrijver:1989:Size,Cygan:2013:Improved, Neuwohner:2021:Improved, Thiery:2023:Improved} for $k$-Set Packing. \\

        However, for weighted $k$-Matroid Intersection the standard local-search algorithm fails even with large swaps and yields a \emph{tight} $k-1$-approximation in the worst-case \cite{Arkin:1998:Local, Lee:2010:Submodular}. Our algorithm instead attempts to apply unweighted algorithms that are known to perform well in a black-box manner.
        It partitions the instance into several weight classes, where each class is an almost unweighted instance. It then performs local-search on each weight class in an iterative fashion. By processing the whole instance greedily over decreasing weight classes, we are able to leverage Lee, Sviridenko and Vondr\'ak's $k/2$-approximation algorithm for the unweighted problem \cite{Lee:2009:Matroid}  and avoid the $k-1$ local-optima. Crucially, this iterative procedure allows us to construct non-improving swaps based on the currently processed weight class. This approach differs from known algorithms which apply various local search methods over the entire instance directly. Beyond our result, our algorithm is a blueprint for solving a weighted maximization problem by decomposing it in many unweighted instances. Using our framework, it seems very likely that an improved local-search algorithm for unweighted $k$-Matroid Intersection will yield an improved approximation for the weighted problem surpassing our guarantees. Note that for unweighted $k$-Set Packing, there is a $(k+1)/3$-approximation algorithm \cite{Cygan:2013:Improved}. We hope that our work will resonate with other problems where weighted variants are challenging.

        \subsection{Construction of matroid exchanges}
            The bulk of our work consists of defining a collection of non-improving swaps between our current solution $A$ and the optimal solution $O$. By processing the instance greedily over decreasing weight classes, we are able to understand the algorithm as follows.
            After processing the $\nth{i}$ weight class, any element of $O$ in the $\nth{i}$ weight class conflicts with the items that were already added to $A$. The optimal elements in the $\nth{i}$ weight class are necessarily conflicting as otherwise we could have added them to $A$ to improve the value of our solution. Thus, each optimal element is either conflicting before or while processing the $\nth{i}$ weight class. We can therefore split the optimal elements contained in the $\nth{i}$ interval into two sets: (1) the elements that conflict within their own weight class, and (2) the elements that conflict with our choices from the previous $i-1$ weight classes. \\

            In matroid terminology, each iteration $i$ defines a large swap between the vertices added to the current solution and conflicting vertices of $O$. Using Rota's exchange property, we refine this swap by constructing two disjoint swaps of smaller size between $A$ and just the vertices from $O$ in the $\nth{i}$ weight-class, and $A$ and just the vertices from $O$ outside the $\nth{i}$ weight-class. The existence of these two swaps allows us in the first case to apply Lee, Sviridenko and Vondr\'ak \cite{Lee:2009:Matroid} analysis and recover their guarantees. This involves another refinement of the large constructed swap that we break this into many small swaps. We point out that our analysis is conceptually different than that for $k$-Set Packing. We start from the existence of a large matroid exchange that we gradually refine to obtain small swaps that our algorithm can find. This top-to-bottom approach diverges from other local-search analyses which typically have a bottom-to-top approach.
            In fact, a fundamental step in the analysis of Berman's algorithm \cite{Berman:2000:d/2} lies in the ability to combine two swaps $(S_1, T_1)$ and $(S_2, T_2)$ to obtain a larger swap $(S_1 \cup S_2, T_1 \cup T_2)$. Berman's analysis \cite{Berman:2000:d/2} heavily relies on this fact to group optimal elements by their \emph{individual} conflicts with edges in $A$ to get a larger swap, and the same is used for \cite{Chandra:1999:Greedy}.
            Unfortunately, for general matroids the union of two swaps is not necessarily a feasible swap. One such example is the graphic matroid on the complete graph with $4$ vertices, $K_4$.
            The techniques used in \cite{Lee:2009:Matroid} are also not directly applicable, as their algorithm can informally be understood as solving $2$-Matroid Intersection and greedily extending the solution. This essentially reduces the number of matroid dependencies by $1$ to obtain a $k-1$-factor.

        \subsection{Computing the approximation ratio}
        The final approximation ratio depends on the following two factors: (1) the fraction of the optimal elements conflicting only within their weight class, and (2) the fraction of optimal elements which conflict with a prior but adjacent weight class.
        As explained, the contribution of (1) can essentially be reduced to the unweighted analysis of Lee, Sviridenko and Vondr\'ak \cite{Lee:2009:Matroid}. Given that the weights within a weight class differ at most by a factor of $(1-\e)$ for some $\e > 0$, we obtain a $\frac{k}{2(1-\e)}$-approximation of the optimal elements conflicting only within their weight class.
        The difficulty then lies in understanding the effect of optimal elements that conflict with a previous weight class.
        This event is \emph{bad} if it behaves like \textsc{Greedy} in that an element in the current solution (from the previous interval) conflicts with up to $k$ optimal elements (in the next interval), all of which have roughly the same weight.
        This suggests that a careful weight class partitioning is needed.
        We use a randomized weight class partitioning to avoid the greedy bottleneck. Instead of choosing the $\nth{i}$ weight class as $[W(1-\e)^{i}, W(1-\e)^{i-1}]$, where $W$ is the largest weight of the instance we randomly shift all intervals by a small factor.
        Using this randomized procedure, we argue that, with high probability, an optimal element is far from being in an anterior weight class. This shows that even if $o$ was conflicting with some element in a prior weight class the weight difference is large enough to improve over the factor $k$ that \textsc{Greedy} attains (with high probability). We achieve an improved approximation guarantee by choosing the parameter $\e$ that balances these two factors.

    \section{Preliminaries}
        For any integer $k \in \NN$, we use the convention $[k] = \{1, \ldots, k\}$.
        We assume that the reader is familiar with basic matroid definitions (for a good survey, see \cite{Schrijver:2003:Combinatorial}). We define the $k$-matroid intersection and matroid $k$-parity problems. All of our results will hold for a more general problem, Matroid $k$-Parity, which is the main focus of this paper. \\

        \textbf{$k$-Matroid Intersection:} Given $k$ matroids $\{\cM_i = (V, \cI_i)\}_{i \in [k]}$ defined on the same groundset $V$ and a linear weight function $w:V \rightarrow \RR_{\geq 0}$, the goal is to find a set $S \subseteq V$ of maximum weight independent in each matroid, i.e. such that $S \in \cI_i$ for all $i \in [k]$. \\

        \textbf{Matroid $k$-Parity:} We are given an edge-weighted hypergraph $G = (V, E)$ on a set of vertices $V$ with linear weight function $w:E\rightarrow \RR_{\geq 0}$ and a matroid $\cM = (V, \cI)$ defined on $V$.
	    Each hyperedge $e \in E$ has size at most $k$.
        The goal is to find a maximum weight collection of disjoint hyperedges $M \subseteq E$ such that the vertices incident to $M$ are independent in $\cI$.
        We assume access to a matroid independence oracle, which for any set $S \subseteq V$ answers whether $S \in \cI$.
        Throughout the document, we will denote by $e(V')$ the set of edges incident to $V' \subseteq V$ with $V' \in \cI$, and conversely let $v(E')$ be the set of vertices contained in $E' \subseteq E$. We say that $A \subseteq E$ is a \emph{feasible solution} if $v(A) \in \cI$.\\

        Observe that if $\cM$ is the \emph{free matroid} (where $\cI = 2^V$), then the problem is known as $k$-Set Packing. As shown by Lee, Sviridenko and Vondr\'ak \cite{Lee:2009:Matroid} finding a maximum weight independent set in the intersection of $k$-matroids can be cast as a matroid $k$-parity problem in which all the edges $e \in E$ are disjoint. In fact without loss of generality, we may assume that the input graph $G = (V, E)$ has the property that each vertex $v$ belongs to a unique hyperedge. Repeating \cite{Lee:2009:Matroid}, we provide a proof of these two facts in \Cref{sec:Reduction} for completeness. As Matroid $k$-Parity generalizes $k$-Matroid Intersection, we provide all algorithms and analyses for the Matroid $k$-Parity.
        It will also be helpful to define a vertex-weighted function.
        \begin{mydef}
            Let $A \subseteq E$ be a collection of disjoint hyperedges, we define $c:v(A) \rightarrow \RR_{\geq 0}$ where $c_v = w_e$ where $v \in e$, and $c(S) = \sum_{v \in S} c_v$ for all $S \subseteq v(A)$. Thus, $c_v$ is the cost of the hyperedge $e\in A$ covering $v$. This function is well defined since $A$ is a collection of disjoint hyperedges.
        \end{mydef}

        Our analyis will use the following two theorems (proved in \Cref{sec:Analysis-SP}) about matroid exchanges.
        \begin{restatable}{mythm}{GeneralizedRota}
            \label{thm:Rota-non-basis-2}
            Let $\cM = (V, \cI)$ be a matroid and $S, T$ be independent sets of $\cM$ with $\card{S} \leq \card{T}$. For any partition $S_1, \ldots, S_m$ of $S$, there are disjoint sets $T_1, \ldots, T_m \subseteq T$ such that $\card{S_i} = \card{T_i}$ and
            $$ S_i \cup (T \bb T_i)  \in \cI.$$
            In particular, if $\card{S} = \card{T}$, then the sets $T_i$ partition $T$.
        \end{restatable}
        This is a slightly more general version of a Rota-exchange where $S$ and $T$ are not necessarily bases of $\cM$. The second proposition, which is an almost immediate corollary, will be useful in our analysis. It states that we can construct exchanges in a top-to-bottom fashion. We can use the existence of a large matroid exchange to repartition into disjoint smaller matroid exchanges.
        \begin{restatable}{mythm}{laminar}
            \label{thm:Laminar-exchanges}
            Let $\cM = (V, \cI)$ be a matroid and $S, T$ be independent sets. Suppose that there is $N_S \subseteq T$ such that  $\card{N_S} = \card{S}$ and $S \cup (T \bb N_S)  \in \cI$. Then, for any partition of $S$ into $S_1, \ldots, S_m$ there is a partition $\{N_{S_i}\}_{i \in [m]}$ of $N_S$ such that $\card{N_{S_i}} = \card{S_i}$ and $S_i \cup (T \bb N_{S_i})  \in \cI$.
        \end{restatable}

    \textbf{Other Notations:} We denote by $W \triangleq \max_{e \in E, v(e) \in \cI} w(e)$ the maximum weight of any feasible edge and denote an interval by $I \subseteq [0, W]$.
        For simplicity, we overload the notation of $I$ and also let $I$ refer to the set of edges $e \in E$ where $w_e \in I$. Thus, given a set of edges $R \subseteq E$, we let $R \cap I \triangleq \lc e \in R \colon w_e \in I \rc$. An interval is then a weight class.
        Finally, given a collection of intervals $\{I_j\}_{j=1}^m$ with $m \in \NN$ and a set of edges $R \subseteq E$, we will denote $R_{\leq i} = \bigcup_{j = 1}^{i} R \cap I_j$, $R_{\geq i} = \bigcup_{j = i}^{m} R \cap I_j$.
    \section{Sliding Local-Search Algorithm}
\label{sec:Algorithm}
    We describe our main algorithm $\SLS$ (Algorithm~\ref{alg:Random-Threshold}), abbreviated by $\SLSS$. $\SLSS$ iteratively runs a local-search algorithm over disjoint intervals of bounded-length in order of decreasing weights.

    \subsection{\ILS}
    $\ILS$ (Algorithm~\ref{alg:local-search}) is a standard local-search algorithm which attempts to improve the current solution by finding improving $(A, I, I)$-\emph{swaps} where $I \subseteq [0, W]$.
    \begin{mydef}[$(A, I, J)$-swap]
        \label{def:feasible-swap}
        Let $A \subseteq E$ be any feasible solution, and $I, J \subseteq [0, W]$ be two intervals. A pair $(S,N)$ is a \emph{$(A, I, J)$-swap} if $S \subseteq (E \bb A) \cap I$ and $\card{S} \leq 2$,  $N \subseteq A \cap J$ and $\card{N} \leq 2k$, and $\left(A \setminus N \right) \cup S$ is a feasible solution. We further say that $(S,T)$ is \emph{improving} if $w(S) > w(N)$.
    \end{mydef}
    We recall that without loss of generality (see \Cref{sec:Disjointness}) all edges of our $k$-Matroid Parity instance are disjoint. Thus, performing an $(A, I, J)$-swap necessarily yields a feasible collection of disjoint hyperedges.

        \begin{algorithm}[H]
        \DontPrintSemicolon
        \caption{\textsc{Interval} \LS$(A, I)$}
        \label{alg:local-search}
        \KwIn{Feasible solution $A \subseteq E$, and interval $I \subseteq [0, W]$}
        \While{There exists $(S, N)$ that is an improving $(A,I, I)$-swap}{
            $A \leftarrow A \bb N \cup S$ \tcp{Improve the solution }
        }
        \Return $A$\;
        \end{algorithm}

    \subsection{\SLS}
    \label{sec:randomized-greedy-description}
    $\SLS$ iteratively calls $\ILS$ on geometrically decreasing bounded length intervals (except the last one).
    Formally, the interval $[0, W]$ is partitioned into $L+1$ disjoint intervals $\{I_j\}_{j = 1}^{L+1}$, for some $L \in \NN$. Each interval $I_j \subseteq [0, W]$ for $j = 1, \ldots, L+1$ is defined by two \emph{markers}. We let $I_j = (m_{j}, m_{j-1}]$ where $m_{j} \in [0, W]$ is the $\nth{j}$-marker.
    The size of each interval is controlled by a parameter $\e > 0$, which will be optimized later, such that $\frac{m_{i-1}}{m_{i}} = 1-\e$.
    Instead of deterministically choosing $m_i = W (1- \e)^{i}$, $\SLSS$ chooses a random threshold $\tau \sim \mathcal{U}[0, \e)$, where $\mathcal{U}[a, b)$ is the uniform distribution between $a$ and $b$, to shift the marker placement and lets $m_i = W (1- \tau)(1-\e)^{i-1}$.
    \begin{mydef}[Markers]
        \label{def:markers}
        Each interval $I_j$ is defined by two \emph{markers} $m_{j-1}$ and $m_j$ such that $I_j = (m_j, m_{j-1}]$ for $j \in [L]$ and $I_{L+1} = [m_{L+1}, m_{L}]$. For some $\tau$ sampled from $\mathcal{U}[0, \e)$, the markers are chosen as:
        $m_{j} = W\cdot (1 -\e)^{j-1}(1-\tau)$ for $j = 0, 1, \ldots, L$ and $m_{L+1} = 0$.
    \end{mydef}

    $\SLSS$ then calls $\ILS$ on $I_1$, then $I_2$ and so on. More precisely, it runs $\ILS$ on $I_1 = (m_1, m_0]$ to find an initial solution $A_{\leq 1} \subseteq I_1$. It then seeks to improve the solution using edges of the next interval. It runs $\ILS(A_{\leq 1}, I_2)$ to obtain solution $A_{\leq 2}\subseteq I_1 \cup I_2$, where $I_2 = (m_2, m_1]$. Since, $\ILS$ attempts to find improving $(A_{\leq 2}, I_2, I_2)$-swap, the edges in $A_{\leq 1} \subseteq I_1$ aren't discarded when processing $I_2$. This implies that $A_{\leq 1} \subseteq A_{\leq 2}$. We denote by $A_{\leq i+1}$ the output of $\ILS(A_{\leq i}, I_{i+1})$. Similarly, we have that $A_{\leq i} \subseteq A_{\leq i+1}$.
    \begin{remark}
        When $\SLSS$ terminates, there is no improving $(A_{\leq i}, I_i, I_i)$-swap for all $i \in [L+1]$.
    \end{remark}

    In \Cref{sec:algorithm-appendix}, we prove, using standard weight-scaling techniques \cite{Lee:2010:Submodular, Lee:2009:Matroid, Berman:2000:d/2} that $\SLSS$  terminates in polynomial time as long as $k$ is constant. We note this is also needed in \cite{Berman:2000:d/2, Cygan:2013:Improved, Lee:2009:Matroid}. Additionally, broadly inspired by streaming algorithms such as \cite{Badanidiyuru-Mirzasoleiman-Karbasi-Krause:2014:Streaming, Kazemi-Mitrovic-Zadimoghaddam-Lattanzi-Karbasi:2019:Submodular}, we prove that we can "discard" edges in the last interval. 

    \begin{restatable}{myprop}{discardOPT}
    \label{thm:discarding-OPT}
        Let $L \triangleq \lceil -\log_{1-\e}(\card{E}\delta^{-1}) \rceil + 1$.
        Let $O$ be the optimal solution and let $A$ be the output of $\SLSS$, and $O_{\leq L} = \{ o \in O \colon w_o \geq m_{L} \}$. Then, $w(O \bb O_{\leq L}) \leq \delta w(O)$.
    \end{restatable}

        \begin{algorithm}[H]
        \DontPrintSemicolon
        \SetKw{KwBy}{by}
        \caption{\SLS$(\delta, \e)$}
        \label{alg:Random-Threshold}
        \KwIn{Parameters $\delta$, $\e$ that quantify the number and the size of the intervals}
        Let $1 - \tau \sim \mathcal{U}(1-\e, 1]$ \hfill \tcp{Random marker placement}
        Let $L \triangleq \lceil -\log_{1-\e}(\card{E}\delta^{-1}) \rceil + 1$ \hfill \tcp{Number of intervals}
        Let $W = \max_{e \in E, v(e) \in \cI} w(e)$\;
        Let $m_0 = W(1-\tau)/(1-\e)$, $m_i = W(1 - \tau)(1-\e)^{i-1}$ for $i \in [L]$, and $m_{L+1} = 0$\;
        \For{$i \gets 0$ \KwTo $L$ \KwBy $1$}
        {$A \gets \textsc{Interval } \LS(A, (m_{i+1}, m_{i}])$\;
        }

        \Return $A$
        \end{algorithm}

\section{A better-than-\texorpdfstring{$k$}{k} approximation guarantee}
    \label{sec: swap construction}
    In this section, we give a simplified analysis of Algorithm~\ref{alg:Random-Threshold} and obtain a $\frac{9}{10}(k+1)$-approximation algorithm. In \Cref{sec:Randomized-Threshold}, we show that the approximation factor can be improved to $(k+1)/(2\ln(2))$. Formally,
    \begin{restatable}{mythm}{mainAPX}\label{thm: apx for k parity}
        For any $k \geq 3$, there is a randomized $\frac{9}{10}(k+1)$-approximation algorithm for weighted Matroid $k$-Parity.
    \end{restatable}
    This is the first result that improves over \textsc{Greedy} which yields a $k$-approximation for Matroid $k$-Parity and over the $k-1$-approximation algorithm of Lee, Sviridenko and Vondr\'ak \cite{Lee:2010:Submodular} for $k$-Matroid Intersection.
    We show in \Cref{sec:high-proba-guarantee} that this guarantee holds with high probability.
    To analyze the approximation guarantee of Algorithm~\ref{alg:Random-Threshold}, we will construct \emph{$(A, I, J)$-swaps} (\Cref{def:feasible-swap}) in \Cref{thm:2nd-layer}. In \Cref{sec: a first layer of exchanges}, we detail our approach, give key definitions and state our main technical lemma (\Cref{thm:2nd-layer}).
    We prove \Cref{thm:2nd-layer} in \Cref{sec:proof-local-swap} using the local optimality of our solution within intervals.
    In \Cref{sec: remaining edges of O}, we analyze the contribution of the swaps across different intervals.

    \subsection{Approach and Definitions}
    \label{sec: a first layer of exchanges}

    To find a collection of local swaps between $A$ and $O$ we think of our algorithm as the following process. Before the start of iteration $i$, our algorithm maintains a solution $v(A_{\leq i-1})$, where $v(A_0) = \emptyset$ and a set $T_{i-1} \subseteq v(O)$ such that $v(A_{\leq i - 1}) \cup (v(O) \bb T_{i-1})$ is independent. The set $T_{i-1}$ are the vertices \emph{conflicting} with $v(A_{i-1})$.
    At the end of iteration $i$, we have extended $v(A_{\leq i-1})$ by finding a set $v(A_i)$. This set further blocks new vertices -- the set $T_i \bb T_{i-1}$.
    \Cref{thm:nested-sets} makes this construction precise.
    This, however, doesn't define local swaps yet. We repeatedly apply Rota's exchange (\Cref{thm:Rota-non-basis-2} and \Cref{thm:Laminar-exchanges}) to define the set of optimal edges that are blocked at any iteration, and then construct local swaps in \Cref{thm:2nd-layer}.
    \begin{restatable}{mylemma}{nestedsets}
        \label{thm:nested-sets}
             There exists a nested sequence of sets $\emptyset = T_0 \subseteq T_1 \subseteq \ldots \subseteq T_{L+1} \subseteq v(O)$ such that $v(A_{\leq i}) \cup (v(O) \bb T_i) \in \cI$
            and $\card{T_i \bb T_{i-1}} = \card{v(A_i)}$.
            Moreover, for any $i \in [L+1]$, and any $o \in O \cap I_i$ such that $w_o > 0$, there exists some $j \leq i$ such that $v(o) \cap T_j \neq \emptyset$.
    \end{restatable}
    \Cref{thm:nested-sets} shows the existence of a set of vertices $T_i \bb T_{i-1} \subseteq v(O)$ which \emph{conflict} with $v(A_i)$ at the $\nth{i}$ iteration and thus have to be discarded. The proof is in \Cref{sec:swaps-appendix}.
    This implies that some optimal edges conflict with the solution $A_{\leq i}$ and we define these edges as follows.
    \begin{mydef}
        Let $O^{(i)} \triangleq \{ o \in O \colon v(o) \cap T_i \neq \emptyset, \mbox{ and } v(o) \cap T_{i-1} = \emptyset\} $ be the edges of $O$ which \emph{conflict} with $A$ for the first time when processing interval $i \in [L+1]$.
    \end{mydef}
    We will construct swaps for edges in $O^{(i)}$. In particular, it will be convenient to expand \Cref{def:feasible-swap} for a partial solution $A_{\leq j}$ at a $\nth{j}$ iteration of the algorithm.
    \begin{mydef}[Local  swap]
        Let $A$ be any feasible solution, and let $(S, N)$ be a pair of sets.
        We say that $(S, N)$ is a \emph{local swap} if there exists $i \in [L+1]$ and $j \leq i$ such that $(S, N)$ is a $(A_{\leq j}, I_i, I_j)$-swap.
    \end{mydef}

    \subsubsection{Recovering the Unweighted Guarantees}\label{sec: singles doubles}
    To leverage the power of local-search in almost unweighted sub-instances, consider the edges of $O^{(i)} \cap I_i$ that lie in $I_i$ and are blocked for the first time at the $\nth{i}$ iteration of Algorithm~\ref{alg:Random-Threshold}. We will prove that these sets behave like an unweighted $k$-matroid parity instance.
    As their weights differ by only a small factor $1-\e$, this implies that the local-search gives a nearly $k/2$-approximation using
    \cite{Lee:2009:Matroid}. For every $i \in [L]$, we define the unweighted portion of the instance as follows,
    \begin{itemize}
    \item $O^{(i)}_s = \lc o \in O^{(i)} \cap I_i \colon \card{v(o) \cap T_i } = 1 \rc$ is the set of blocked edges in the $\nth{i}$ interval with a \emph{single} vertex conflict, and let $ O_s = \bigcup_{i \in [L]} O^{(i)}_s$ be the set of \emph{singles}.
    \item $O^{(i)}_d = \lc o \in O^{(i)} \cap I_i \colon \card{v(o) \cap T_i } \geq 2 \rc$ is the set of blocked edges in the $\nth{i}$ interval with a \emph{double} vertex conflict, and let $O_d = \bigcup_{i \in [L]} O^{(i)}_d$ be the set of \emph{doubles}.
    \end{itemize}
    Note that these sets are only defined up to the $\nth{L}$ interval.
    Equipped with the above definitions, we state our main result that constructs local swaps $\{(o, e(N_o))\}_{o \in O}$ between $A$ and $O$ at each iteration of Algorithm~\ref{alg:Random-Threshold}.
    Each swap is non-improving and satisfies a nice partition property such that each edge $a \in A$ isn't part of more than $k$ swaps.
    Additionally, every $a \in  A$ appears in the local swap of at most 1 \emph{single} optimal edge.
    We also show that for every $o \in O_d$, its weight is much smaller than the weight of its neighborhood.
    This will allow us to obtain two straightforward corollaries. The first one says that Algorithm~\ref{alg:Random-Threshold} is always a $k$-approximation. The second bounds the total weight of $O_s$ (similarly to \cite{Lee:2009:Matroid}).
    \begin{mylemma}
        \label{thm:2nd-layer}
        For any random choice of $\tau$, there exists a collection $\{(o, e(N_o))\}_{o \in O}$ of local swaps such that
        \begin{enumerate}[label=\textcolor{red}{\arabic*.}, ref=\arabic*]
            \item for all $i \in [L+1]$, and every $o \in O^{(i)}$, we have $\emptyset \neq N_o \subseteq v(A_i)$ and $w_o \leq c(N_o)$, \label{enum:1b}
            \item each $v \in v(A)$ appears in at most one of the sets $\{N_o\}_{o \in O}$, \label{enum:2b}
            \item for every $o \in O_s$, $c(N_o) = w(e(N_o))$ and for $o' \neq o \in O_s$, we have $e(N_o) \cap e(N_o') = \emptyset$, \label{enum:3b}
            \item for every $ o \in O_d$, we have that $\card{N_o} \geq 2$ and $w_o \leq \frac{1}{2(1-\e)} c(N_o)$. \label{enum: 4b}
        \end{enumerate}
    \end{mylemma}

    \begin{mycor}
    \label{cor: k approximation factor}
        Let  $\{(o, e(N_o))\}_{o \in O}$ be the collection from \Cref{thm:2nd-layer}, then
        $\displaystyle\sum_{o \in O} c(N_o) \leq k w(A)$. In particular, we have that $w(O) \leq k w(A)$.
    \end{mycor}
    \begin{proof}[Proof of \Cref{cor: k approximation factor}]
        By \Cref{thm:2nd-layer}, each $v \in v(A)$ appears at most once in one set $N_o$ for $o \in O$. Since every $a \in A$ contains at most $k$ vertices of $v(A)$ we get that,
        \begin{align*}
            \displaystyle\sum_{o \in O} c(N_o)
            & = \displaystyle\sum_{a \in A} w_a \displaystyle\sum_{o \in O} \card{v(a) \cap N_o}
            \leq  \displaystyle\sum_{a \in A} w_a \cdot k
            =  k \cdot w(A). \qedhere
        \end{align*}
        Applying Property~\ref{enum:1b} of \Cref{thm:2nd-layer} and the previous bound, we get that:
        $w(O) \leq \sum_{o \in O} c(N_o) \leq k w(A). $
    \end{proof}

    \begin{mycor}\label{cor: weight Os at most w(A)}
        By the properties~\ref{enum:1b} and \ref{enum:3b} of \Cref{thm:2nd-layer}, we have $w(O_s) \leq w(A)$.
    \end{mycor}

    \subsection{Proof of \texorpdfstring{\Cref{thm:2nd-layer}}{thm}}
    \label{sec:proof-local-swap}
    \begin{proof}[Proof of \Cref{thm:2nd-layer}]
        Let $\{T_i\}_{i=1}^{L+1}$ be the sequence of sets obtained from \Cref{thm:nested-sets}. We apply Rota exchange properties (\Cref{thm:Rota-non-basis-2}) on $\cM_i$, the matroid contracted on $v(A_{\leq i-1}) \cup (v(O) \bb T_i)$, to obtain local swaps $\{(o, e(N_o))\}_{o \in O}$. By \Cref{thm:nested-sets}, the sets $T_i \bb T_{i-1}$ and $v(A_i)$ are independent in $\cM_i$ and $\card{v(A_i)} = \card{T_i \bb T_{i-1}}$.
        We partition $T_i \bb T_{i-1}$ as follows: let $B_s = v(O^{(i)}_s) \cap T_i$ and $B' = v(O^{(i)} \bb O^{(i)}_s) \cap T_i$ and $B'' = (T_i \bb T_{i-1}) \bb (B_s \cup B')$. Note that this is a partition of $T_i \bb T_{i-1}$ by definition of $O^{(i)}$. Here $B_s$ is simply empty when $i = L+1$.
        By \Cref{thm:Rota-non-basis-2}, there exist disjoint sets $N_s, N' \subseteq v(A_i)$ such that $v(A_{i}) \bb N_s \cup B_s$ and $v(A_{i}) \bb N' \cup B'$ are independent in $\cM_i$ and $\card{N_s} = \card{B_s}$, and $\card{N'} = \card{B'}$ respectively.
        Property~\ref{enum:3b} follows from the following two claims which we now prove. Let $i \in [L]$, then
        \begin{enumerate}
            \item For all $a \in A_{i}$, we have that $\card{v(a) \cap N_s} \leq 1$,
            \item $N_s = \bigsqcup_{o \in O_s^{(i)}} N_o$, and $\card{N_o} = 1$ for $o \in O^{(i)}_s$.
        \end{enumerate}

        We start with the first claim. Suppose conversely that there exists an edge $a \in A_i$ such that $\card{v(a) \cap N_s} \geq 2$. This implies that $|v(A_i) \bb (N_s \cup v(a)) \cup B_s| = |v(A_i)| - (|N_s| + |v(a)| - |N_s \cap v(a)|) + |B_s| \geq \card{v(A_i - a)} + 2$ since $|N_s| = |B_s|$.
        Because $v(A_i) \bb (N_s \cup v(a)) \cup B_s$ and $v(A_i - a)$ are independent in $\cM_i$, we can find an extension ${b, b'} \in B_s$ such that $v(A_i - a) \cup \{b, b'\} \in \cI_i$. Observe that the vertices $b,b'$ belong to distinct hyperedges in $O^{(i)}_s$, which we denote by $e, e' \in O^{(i)}_s$, respectively.
        Since $v(A_i - a) \cup \{b, b'\}$ is independent in $\cM_i$, we have that $(v(A_{\leq i} - a) \cup \{b, b'\} \cup (v(O) \bb T_{i}) \in \cI$. In particular, $A_{\leq i} \bb \{a\} \cup \{e, e'\}$ is feasible since $\{e, e'\} \subseteq \{b, b'\} \cup (v(O) \bb T_{i})$ by definition of $O^{(i)}$.
        We prove that this yields an improving $(A_{\leq i}, I_i)$-swap as long as $\e \in (0, 0.5)$ and $i \in [L]$. We have that $a \in A_i \subseteq I_i$ and $e, e' \in O^{(i)}_s \subseteq I_i$. Since $I_i = (m_i, m_{i-1}]$, we have
        \begin{align*}
            w_e + w_e' - w_a & > 2m_i - m_{i-1}
            =   2(1-\e)m_{i-1} - m_{i-1}
             =  m_{i-1}(1 - 2\e)
            > 0,
        \end{align*}
        where we use $m_i = (1-\e) m_{i-1}$ for $i \in [L]$,
        contradicting the absence of improving $(A_{\leq i}, I_i)$-swaps. \\

        We prove the next claim together with Property~\ref{enum:1b} and \ref{enum:2b}.
        For every $i \in [L+1]$, we partition $B_s$ into $\{ v(o) \cap B_s \}_{o \in O^{(i)}_s}$ and $B'$ into $\{ v(o) \cap B' \}_{o \in O^{(i)} \bb O^{(i)}_s}$. 
        Applying \Cref{thm:Laminar-exchanges}, we get that there are collections of disjoints sets $\{N_o\}_{o \in O^{(i)}_s}$ and $\{N'_o \}_{o \in O^{(i)} \bb O^{(i)}_s}$ partitioning $N_s$ and $N'$ such that $v(A_i) \bb N_o \cup (v(o) \cap B_s) \in \cI_i$ for all $o \in O^{(i)}_s$ and $v(A_i) \bb N'_o \cup (v(o) \cap B') \in \cI_i$ for all $o \in O^{(i)} \bb O^{(i)}_s$, respectively. \Cref{thm:Laminar-exchanges} further says that $\card{N_o'} = \card{v(o) \cap B'} \geq 1$ and $\card{N_o} = \card{v(o) \cap B_s} = 1$, where the second equality uses that $\card{v(o) \cap T_i} = 1$ for all $o \in O^{(i)}_s$.
        We prove that $\{(o, e(N_o))\}_{o \in O^{(i)}_s}$ and $\{(o, e(N'_o))\}_{o \in O^{(i)}\bb O^{(i)}_s}$ are local swaps. Fix $i \in [L+1]$ and let $o \in O^{(i)}_s$ together with its pair $(o, e(N_o))$. The proof is identical for $o \in O^{(i)} \bb O^{(i)}_s$. We prove that the swap is independent:
        \begin{align*}
            v(A_{\leq i}) \bb \{v(e(N_o))\} \cup v(o) & \subseteq v(A_{\leq i}) \bb N_o \cup v(o) \\
            & \subseteq \Big( v(A_{i}) \bb N_o \cup (v(o) \cap T_i) \Big)  \cup \Big( v(A_{\leq i-1}) \cup (v(O) \bb T_i) \Big)\\
            & \subseteq \Big( v(A_{i}) \bb N_o \cup (v(o) \cap (T_i \bb T_{i-1})) \Big)  \cup \Big( v(A_{\leq i-1}) \cup (v(O) \bb T_i) \Big)
        \end{align*}
        where we used in the last containment that $v(o) \cap T_{i-1} = \emptyset$ for all $o \in O^{(i)}$. The last term is independent since $v(o) \cap B_s = v(o) \cap T_{i} \bb T_{i-1}$ and $v(A_{i}) \bb N_o \cup (v(o) \cap B_s)$ is independent in $\cM_i$. We have that $(o, e(N_o))$ is, in fact, a $(A_{\leq i}, I_\ell, I_i)$-swap for some $\ell \geq i$, since $o \in I_\ell$ by the second part of \Cref{thm:nested-sets}.
        We now prove Property~\ref{enum:1b}. Let $v \in N_o$ (which exists since $\card{N_o} \geq 1$) and let $a$ be the edge incident to $v$.
        If $i = \ell$, then the property follows from the absence of $(A_{\leq i}, I_i)$-swap, which implies that $w_o \leq w(e(N_o)) \leq c(N_o)$.
        Then, suppose $i < \ell$, we have that $w_o \leq m_{\ell - 1} \leq m_i < w_a \leq c(N_o)$ by definition of the markers. Property~\ref{enum:2b} follows from the fact that that $\{v(A_i)\}_{i \in [L+1]}$ are disjoint sets of vertices and that the sets $N_o$ are also disjoint.\\

        Lastly, we show these local swaps satisfy Property \ref{enum: 4b} for $o \in O_d$.
        Let $i \in [L]$ be any index and fix $o \in O^{(i)}_d$. By construction of the local swaps $(o, e(N'_o))$, we have $\card{N'_o} = \card{v(o) \cap B'} = \card{v(o) \cap T_i \bb T_{i-1}} \geq 2$.
        Let $v_1, v_2 \in N_o$ be two distinct vertices .
        Further, let $a_1, a_2 \subseteq e(N_o) \subseteq A_i$ be the edges incident to $v_1$ and $v_2$. It is possible that $a_1 = a_2$.
        Since $a_j \in A_i$, we have that $m_{i-1} \geq w_{a_j} > m_{i}$ for $j = 1, 2$.
        Additionally, we have that $o \in O^{(i)}_d$, so $o$ must lie in $I_i$. This means that $w_o \leq m_{i-1}$.
            Therefore,
            \begin{align*}
                w_o & = \frac{w_o}{c(v_1) + c(v_2)} \lb c(v_1) + c(v_2) \rb \leq  \frac{m_{i-1}}{2m_{i}} c(N_o)
                = \frac{1}{2(1-\e)} c(N_o). \qedhere
            \end{align*}
    \end{proof}

    \subsection{Escaping local-optima using randomness}
    \label{sec: remaining edges of O}
    By Property~\ref{enum:3b} and \ref{enum: 4b} of \Cref{thm:2nd-layer} we are able to analyze the absence of improving swap in a given interval. Unfortunately, the optimal edges in the $\nth{i}$ interval might already be blocked after processing an anterior interval $j < i$. This poses problems as an edge in $a \in A_{i-1}$ may block up to $k$ edges in $O \cap I_i$ that differ in weight only slightly. If this happens, then $\SLSS$ has the same guarantees as \textsc{Greedy}. Importantly, this situation can be avoided by slightly shifting the $\nth{i}$-marker. We use the random marker placement to show that such events occur with low probability.
    This motivates the following definitions.
    Let $\gamma \in (0,1)$ be a fixed constant.
    \begin{mydef}
        \label{def:closest-marker}
        For each $o \in O$, we define by $m_o$ the closest marker to $w_o$ such that $m_o \geq w_o$.
    \end{mydef}
    We partition the remaining edges of $O \bb (O_s \cup O_d)$ into those that are close to $m_o$ and 
    \begin{itemize}
        \item $O_b = \lc o \in O \bb (O_s \cup O_d): w_o \geq (1 + \gamma)^{-1} m_{o} \rc$ is the set of blocked edges whose weight is close to $m_o$.
        These edges are \emph{bad} for the analysis.
        \item $O_g = \lc o \in O \bb (O_s \cup O_d): w_o < (1 + \gamma)^{-1} m_{o}\rc$ is the set of blocked edges whose weight is far from $m_o$.
        These edges are \emph{good} for the analysis as shown in \Cref{thm:good-improve} proved in \Cref{sec:good-improve-appendix}.
    \end{itemize}
    \begin{restatable}{mylemma}{goodimprove}
        \label{thm:good-improve}
        Let $o \in O_g \cap O_{\leq L}$ and $(o, e(N_o))$ be its local swap  given by \Cref{thm:2nd-layer}. Then,
        \begin{align*}
            (1+\gamma) w_o & \leq c(N_o).
        \end{align*}
    \end{restatable}

    Motivated by \Cref{thm:good-improve}, our objective is then to estimate the expected weight of $O_b$. In \Cref{thm:delta-close}, we use the random threshold to argue that the expected weight is small.  

    \begin{restatable}{mylemma}{deltaclose}
        \label{thm:delta-close}
        Let $\e > 0$ be the parameter from Algorithm~\ref{alg:Random-Threshold}. For any $o \in O$ and  $ \gamma \in \ld 0, (1 - \e)^{-1} -1 \rb $, we have that
        \begin{align*}
            \probtau{o \in O_b} \leq \probtau{w_o \geq (1+\gamma)^{-1} m_o} \leq \frac{\gamma}{\e(1+\gamma)}.
        \end{align*}
        In particular, the following holds: $\esp{w(O_b)} \leq \frac{\gamma}{\e(1+\gamma)} w(O)$.
    \end{restatable}
    The first inequality follows from the definition of $O_b$. Thus, \Cref{thm:delta-close} tells that it is sufficient to compute the probability that an edge $o$ lands close to $m_o$ to bound the contribution of $O_b$.
    \begin{proof}[Proof of \Cref{thm:delta-close}]
    Fix any $o \in O$. We start by proving the lemma when $w_o \in [1 - \e, 1]\cdot W$ and split the proof into two computations depending on the value of $w_o$.
        Suppose that $w_o \in [ 1- \e, \frac{1}{1+\gamma} ]\cdot W$. Then, the only marker for which $w_o \geq (1+\gamma)^{-1} m_o$ is if $m_o = m_1$. Thus,
    \begin{align*}
        \probtau{w_o \geq (1+\gamma)^{-1} m_o} &
        = \probtau{m_1 \in [w_o, w_o (1 + \gamma)]}
        = \frac{1}{\e \cdot W} \cdot \lb w_o( 1 + \gamma) - w_o \rb
        \leq \frac{\gamma}{\e(1 + \gamma)}.
    \end{align*}
    In the second inequality, we used that $m_1 = (1 - \tau)W \sim \mathcal{U}(1 - \e, 1]\cdot W$ and that $[w_o, w_o (1 + \gamma)] \subseteq [1 - \e, 1]\cdot W$.
    The last inequality uses that $w_o \leq \frac{W}{1+\gamma}$.
    Suppose alternatively that $w_o \in [ \frac{1}{1+\gamma}, 1]\cdot W$. Then, we have:
    \begin{align*}
        \probtau{w_o \geq (1+\gamma)^{-1} m_o}
        & = \probtau{ \lb m_1 \in [w_o, 1] \rb  \vee \lb m_0 \in [1, (1 + \gamma)w_o] \rb} \\
        & = \probtau{ m_1 \in [w_o, 1]} + \probtau{ m_1 \in [1 - \e, (1-\e)(1 + \gamma)w_o]} \\
        & =  \frac{1}{\e \cdot W} \cdot \lb \e + w_o \lb (1 - \e)( 1+ \gamma) - 1 \rb \rb \\
        & \leq \frac{1}{\e} \cdot \frac{\gamma}{1 + \gamma}.
    \end{align*}
        In the second equality, we use that events are disjoint from each other and that $m_0( 1- \e) = m_1$. The third equality uses that $m_1 \sim \mathcal{U}(1 - \e, 1]\cdot W$. In the last inequality, we observe that $1 + \gamma \leq (1 - \e)^{-1}$ so $(1 - \e)( 1+ \gamma) - 1 \leq 0$. Substituting $w_o \geq (1 +\gamma)^{-1} W$ into the last inequality yields the desired result. This proves the result when $w_o \in [1- \e, 1]\cdot W$.

        Suppose now that $w_o \in [(1- \e)^{i}, (1- \e)^{i-1}]\cdot W$ for some $i \in [L+1]$, we reduce this case to the previous computation. Indeed, the probability that $w_o \in [\frac{1}{1+\gamma}m_j, m_j]$ for some $j$ is equal to the probability that the point $\frac{w_o}{(1 - \e)^{i-1}}$ is in $[\frac{1}{(1+\gamma)} m_{j-i +1}, m_{j-i+1}]$. Note that $\frac{w_o}{(1 - \e)^{i-1}} \in [1- \e, 1]\cdot W$, thus, by the previous computation, we get that $o$ is in $O_b$ with probability at most $\frac{\gamma}{\e(1+\gamma)}$. Lastly, suppose that $w_o \in [0, (1-\e)^{L+1}]\cdot W$. Then as $m_L = W(1-\tau)(1-\e)^{L-1}$, we have that $(1+\gamma)^{-1}m_o = (1+\gamma)^{-1}m_L \geq W (1-\e)^{L+1} \geq w_o$, where we used that $(1-\tau) \geq 1-\e$ and $(1+\gamma)^{-1} \geq 1 - \e$. This implies that $o \in O_g$.
    \end{proof}

    \subsection{Obtaining an \texorpdfstring{$\Omega(k)$}{omega k} approximation ratio}
    \label{sec:APX-ratio}
    We combine the improvements due to the local-search analysis of \Cref{thm:2nd-layer} and to  the random marker analysis of \Cref{thm:delta-close} to show that Algorithm~\ref{alg:Random-Threshold} is an $\Omega(k)$-approximation algorithm for weighted $k$-Matroid Intersection and weighted $k$-Matroid Parity. The full proof is in \Cref{sec:final-approx-value-apdx}.
    \begin{restatable}{mythm}{APXsimple}
        \label{thm:APX-easy}
        Let $\e \in (0, 1/2)$, $\delta \in (0, 1)$ be the parameters of Algorithm~\ref{alg:Random-Threshold}, and let $\gamma \in (0, 1)$ be a constant such that $\gamma \leq (1-\e)^{-1} - 1$ and $\gamma \leq 2(1-\e) - 1$.
        Let $A$ be the output of Algorithm~\ref{alg:Random-Threshold}. Then,
        \begin{align*}
            w(O) & \leq \frac{k + \gamma}{1+ \gamma \lb 1- \frac{\gamma}{\e(1+\gamma)} \rb - \delta } \cdot \esp{w(A)}.
        \end{align*}
        Setting $\e = 0.3873$, $\gamma = 0.2253$ and $\delta$ small enough we get that
        $w(O) \leq \frac{9(k + 1)}{10} \cdot \esp{w(A)}$.
        \end{restatable}

        \subsection{Conclusion and Open Questions}
\label{sec:Conclusion}
    We study weighted $k$-matroid intersection and its generalization as weighted matroid $k$-parity. In contrast to special cases, such as $k$-set packing, both problems are not as well understood. Prior to this work, \textsc{Greedy} was the state-of-the-art algorithm with a $k$-approximation. We significantly improve over this bound and provide a $\frac{k+1}{2\ln(2)} \simeq 0.722(k+1)$-approximation which largely reduces the gap between the unweighted and weighted setting.
    Our algorithm is conceptually simple and provides an alternative to the current algorithms for special cases such as Berman's algorithm. We iteratively apply a weighted local-search over random intervals which prevents the algorithm from reaching certain bad local optima. Beyond an improved constant, our algorithm offers a blueprint for transforming algorithms over the unweighted setting into algorithms over the weighted setting. This may be of independent interest and could lead to improvements for other problems beyond Matroid $k$-Parity.

    \paragraph{Open Questions:}
        It remains interesting to close the gap between the unweighted and the weighted setting, even in the simpler case of $k$-Set Packing for which there is still a gap \cite{Cygan:2013:Improved, Neuwohner:2023:Passing}. On the hardness front, as mentioned in \cite{Lee:2024:Asymptotically}, better hardness of approximation for $k$-CSP might improve the $k/12$-hardness of $k$-Dimensional Matching which would extend to $k$-Matroid Intersection. Finally, a clear direction for future improvements is to investigate whether there exists a $\frac{k+1}{3}$-approximation for unweighted $k$-matroid intersection similarly to unweighted $k$-set packing or give a separation.


\subsection*{Acknowledgments}
The authors are grateful to Ola Svensson for the many helpful discussions and comments.

\bibliography{biblio}

\newcommand{\etalchar}[1]{$^{#1}$}
\begin{thebibliography}{BMKK14}

\bibitem[AH98]{Arkin:1998:Local}
Esther~M. Arkin and Refael Hassin.
\newblock On local search for weighted \emph{k}-set packing.
\newblock {\em Math. Oper. Res.}, 23(3):640--648, 1998.

\bibitem[Ber00]{Berman:2000:d/2}
Piotr Berman.
\newblock A d/2 approximation for maximum weight independent set in d-claw free graphs.
\newblock In {\em Scandinavian Workshop on Algorithm Theory}, pages 214--219. Springer, 2000.

\bibitem[BK03]{Berman:2003:Optimizing}
Piotr Berman and Piotr Krysta.
\newblock Optimizing misdirection.
\newblock In {\em Proceedings of the Fourteenth Annual {ACM-SIAM} Symposium on Discrete Algorithms, January 12-14, 2003, Baltimore, Maryland, {USA}}, pages 192--201. {ACM/SIAM}, 2003.

\bibitem[BMKK14]{Badanidiyuru-Mirzasoleiman-Karbasi-Krause:2014:Streaming}
Ashwinkumar Badanidiyuru, Baharan Mirzasoleiman, Amin Karbasi, and Andreas Krause.
\newblock Streaming submodular maximization: massive data summarization on the fly.
\newblock In Sofus~A. Macskassy, Claudia Perlich, Jure Leskovec, Wei Wang, and Rayid Ghani, editors, {\em The 20th {ACM} {SIGKDD} International Conference on Knowledge Discovery and Data Mining, {KDD} '14, New York, NY, {USA} - August 24 - 27, 2014}, pages 671--680. {ACM}, 2014.

\bibitem[CH99]{Chandra:1999:Greedy}
Barun Chandra and Magn{\'{u}}s~M. Halld{\'{o}}rsson.
\newblock Greedy local improvement and weighted set packing approximation.
\newblock In Robert~Endre Tarjan and Tandy~J. Warnow, editors, {\em Proceedings of the Tenth Annual {ACM-SIAM} Symposium on Discrete Algorithms, 17-19 January 1999, Baltimore, Maryland, {USA}}, pages 169--176. {ACM/SIAM}, 1999.

\bibitem[CLL14]{Cheung-Lau-Leung:2014:Algebraic}
Ho~Yee Cheung, Lap~Chi Lau, and Kai~Man Leung.
\newblock Algebraic algorithms for linear matroid parity problems.
\newblock {\em ACM Transactions on Algorithms (TALG)}, 10(3):1--26, 2014.

\bibitem[Cyg13]{Cygan:2013:Improved}
Marek Cygan.
\newblock Improved approximation for 3-dimensional matching via bounded pathwidth local search.
\newblock In {\em 54th Annual {IEEE} Symposium on Foundations of Computer Science, {FOCS} 2013, 26-29 October, 2013, Berkeley, CA, {USA}}, pages 509--518. {IEEE} Computer Society, 2013.

\bibitem[Edm03]{Edmonds:2003:Submodular}
Jack Edmonds.
\newblock {\em Submodular Functions, Matroids, and Certain Polyhedra}, pages 11--26.
\newblock Springer Berlin Heidelberg, Berlin, Heidelberg, 2003.

\bibitem[Gal68]{Gale:1968:Optimal}
David Gale.
\newblock Optimal assignments in an ordered set: an application of matroid theory.
\newblock {\em Journal of Combinatorial Theory}, 4(2):176--180, 1968.

\bibitem[HS89]{Hurkens-Schrijver:1989:Size}
Cor A.~J. Hurkens and Alexander Schrijver.
\newblock On the size of systems of sets every t of which have an sdr, with an application to the worst-case ratio of heuristics for packing problems.
\newblock {\em {SIAM} J. Discret. Math.}, 2(1):68--72, 1989.

\bibitem[HSS06]{Hazan-Safra-Schwartz:2006:Complexity}
Elad Hazan, Shmuel Safra, and Oded Schwartz.
\newblock On the complexity of approximating \emph{k}-set packing.
\newblock {\em Comput. Complex.}, 15(1):20--39, 2006.

\bibitem[IK17]{Iwata-Kobayashi:2017:Weighted}
Satoru Iwata and Yusuke Kobayashi.
\newblock A weighted linear matroid parity algorithm.
\newblock In Hamed Hatami, Pierre McKenzie, and Valerie King, editors, {\em Proceedings of the 49th Annual {ACM} {SIGACT} Symposium on Theory of Computing, {STOC} 2017, Montreal, QC, Canada, June 19-23, 2017}, pages 264--276. {ACM}, 2017.

\bibitem[JK82]{Jensen-Korte:1982:Complexity}
Per~M. Jensen and Bernhard Korte.
\newblock Complexity of matroid property algorithms.
\newblock {\em {SIAM} J. Comput.}, 11(1):184--190, 1982.

\bibitem[KMZ{\etalchar{+}}19]{Kazemi-Mitrovic-Zadimoghaddam-Lattanzi-Karbasi:2019:Submodular}
Ehsan Kazemi, Marko Mitrovic, Morteza Zadimoghaddam, Silvio Lattanzi, and Amin Karbasi.
\newblock Submodular streaming in all its glory: Tight approximation, minimum memory and low adaptive complexity.
\newblock In {\em International Conference on Machine Learning}, pages 3311--3320. PMLR, 2019.

\bibitem[Las15]{Lason:2015:List-Coloring}
Michal Lason.
\newblock List coloring of matroids and base exchange properties.
\newblock {\em Eur. J. Comb.}, 49:265--268, 2015.

\bibitem[Law01]{Lawler:2001:Combinatorial}
Eugene~L Lawler.
\newblock {\em Combinatorial optimization: networks and matroids}.
\newblock Courier Corporation, 2001.

\bibitem[Lov80]{Lovasz:1980:Matroid-Matching}
L{\'{a}}szl{\'{o}} Lov{\'{a}}sz.
\newblock Matroid matching and some applications.
\newblock {\em J. Comb. Theory {B}}, 28(2):208--236, 1980.

\bibitem[LST24]{Lee:2024:Asymptotically}
Euiwoong Lee, Ola Svensson, and Theophile Thiery.
\newblock Asymptotically optimal hardness for $ k $-set packing and $ k $-matroid intersection.
\newblock {\em arXiv preprint arXiv:2409.17831}, 2024.

\bibitem[LSV09]{Lee:2010:Submodular}
Jon Lee, Maxim Sviridenko, and Jan Vondr{\'{a}}k.
\newblock Submodular maximization over multiple matroids via generalized exchange properties.
\newblock In Irit Dinur, Klaus Jansen, Joseph Naor, and Jos{\'{e}} D.~P. Rolim, editors, {\em Approximation, Randomization, and Combinatorial Optimization. Algorithms and Techniques, 12th International Workshop, {APPROX} 2009, and 13th International Workshop, {RANDOM} 2009, Berkeley, CA, USA, August 21-23, 2009. Proceedings}, volume 5687 of {\em Lecture Notes in Computer Science}, pages 244--257. Springer, 2009.

\bibitem[LSV10]{Lee:2009:Matroid}
Jon Lee, Maxim Sviridenko, and Jan Vondr{\'{a}}k.
\newblock Matroid matching: the power of local search.
\newblock In Leonard~J. Schulman, editor, {\em Proceedings of the 42nd {ACM} Symposium on Theory of Computing, {STOC} 2010, Cambridge, Massachusetts, USA, 5-8 June 2010}, pages 369--378. {ACM}, 2010.

\bibitem[Neu21]{Neuwohner:2021:Improved}
Meike Neuwohner.
\newblock An improved approximation algorithm for the maximum weight independent set problem in d-claw free graphs.
\newblock In Markus Bl{\"{a}}ser and Benjamin Monmege, editors, {\em 38th International Symposium on Theoretical Aspects of Computer Science, {STACS} 2021, March 16-19, 2021, Saarbr{\"{u}}cken, Germany (Virtual Conference)}, volume 187 of {\em LIPIcs}, pages 53:1--53:20. Schloss Dagstuhl - Leibniz-Zentrum f{\"{u}}r Informatik, 2021.

\bibitem[Neu22]{Neuwohner:2022:Limits}
Meike Neuwohner.
\newblock The limits of local search for weighted k-set packing.
\newblock In Karen~I. Aardal and Laura Sanit{\`{a}}, editors, {\em Integer Programming and Combinatorial Optimization - 23rd International Conference, {IPCO} 2022, Eindhoven, The Netherlands, June 27-29, 2022, Proceedings}, volume 13265 of {\em Lecture Notes in Computer Science}, pages 415--428. Springer, 2022.

\bibitem[Neu23]{Neuwohner:2023:Passing}
Meike Neuwohner.
\newblock Passing the limits of pure local search for weighted \emph{k}-set packing.
\newblock In Nikhil Bansal and Viswanath Nagarajan, editors, {\em Proceedings of the 2023 {ACM-SIAM} Symposium on Discrete Algorithms, {SODA} 2023, Florence, Italy, January 22-25, 2023}, pages 1090--1137. {SIAM}, 2023.

\bibitem[Rad57]{Rado:1957:Note}
Richard Rado.
\newblock Note on independence functions.
\newblock {\em Proceedings of the London Mathematical Society}, 3(1):300--320, 1957.

\bibitem[S{\etalchar{+}}03]{Schrijver:2003:Combinatorial}
Alexander Schrijver et~al.
\newblock {\em Combinatorial optimization: polyhedra and efficiency}, volume~24.
\newblock Springer, 2003.

\bibitem[TW23]{Thiery:2023:Improved}
Theophile Thiery and Justin Ward.
\newblock An improved approximation for maximum weighted \emph{k}-set packing.
\newblock In Nikhil Bansal and Viswanath Nagarajan, editors, {\em Proceedings of the 2023 {ACM-SIAM} Symposium on Discrete Algorithms, {SODA} 2023, Florence, Italy, January 22-25, 2023}, pages 1138--1162. {SIAM}, 2023.

\end{thebibliography}
\bibliographystyle{alpha}
\newpage

\appendix


    \section{Omitted Proofs}
\label{sec:Supplementary Material}
    For completeness, we provide the missing proofs from \Cref{sec: swap construction}.
    \subsection{Proof of \texorpdfstring{\Cref{thm:nested-sets}}{thm}}
    \label{sec:swaps-appendix}
        \nestedsets*
            Without loss of generality, we may assume that $\card{v(O)} \geq \card{v(A)}$ by adding disjoint edges of weight equal to $0$, each incident to $k$ new dummy vertices independent with everything else.
            We prove the statement by induction on $i$. Let $v(A_1)$ be the vertices incident to $A_1$ and apply \Cref{thm:Rota-non-basis-2} to find a set $T_1 \subseteq v(O)$ such that
            $ v(A_1) \cup (v(O) \bb T_1) \in \cI$ and $\card{T_1} = \card{v(A_1)}$. This concludes the proof for $i = 1$.\\

            By the induction hypothesis, we have a collection $T_0 \subseteq \ldots \subseteq T_{i-1} \subseteq v(O)$ such that $v(A_{\leq i-1}) \cup (v(O) \bb T_{i-1}) \in \cI$ and $\card{T_{i-1} \bb T_{i-2}} = \card{v(A_{i-1})}$. Let $\cM' = (V, \cI')$ be the matroid $\cM$ contracted on $v(A_{\leq i-1})$. Observe that $v(O) \bb T_{i-1}$ and $v(A_{i})$ are independent in $\cM'$.
            Moreover, by the induction hypothesis $ 0 \geq |v(A_{\leq i})| - |v(O)| =  |v(A_{\leq i-1})| + |v(A_i)| - |v(O) \bb T_{i-1}| - |T_{i-1}| = |v(A_i)| - |v(O) \bb T_{i-1}|$. So $|v(O) \bb T_{i-1}| \geq |v(A_i)|$.
            Thus, applying \Cref{thm:Rota-non-basis-2} to $v(A_i)$ and $v(O) \bb T_{i-1}$ in $\cM'$, we get that there exists a set $T_i \bb T_{i-1} \subseteq v(O) \bb T_{i-1}$ such that $v(A_i) \cup (v(O) \bb T_{i-1}) \bb (T_i\bb T_{i-1}) \in \cI'$ and that $\card{T_i \bb T_{i-1}} = \card{v(A_i)}$. We finish the induction by uncontracting the matroid.\\

            It remains to prove the second part of the statement. Let $o \in O \cap I_i$ be such that $w_o > 0$ for some $i \in [L+1]$. Suppose by contradiction that $v(o) \cap T_j = \emptyset$ for all $j \leq i$. This implies that:
            $$ v(A_{\leq i}) \cup v(o) \subseteq v(A_{\leq i}) \cup v(o) \cup \lb v(O) \bb T_{i} \rb = v(A_{\leq i}) \cup \lb v(O) \bb T_{i} \rb \in \cI, $$
            where we used the first part of the lemma to claim the independence of the right-hand side. So $v(o)$ can be added to $v(A_{\leq i})$ without violating the independence constraint. Since $w_o > 0 $, the pair $(\{o\}, \emptyset)$ is an improving $(A_{\leq i}, I_i, I_i)$-swap contradicting the termination of Algorithm~\ref{alg:Random-Threshold}.

    \subsection{Proof of \texorpdfstring{\Cref{thm:good-improve}}{thm}}
    \label{sec:good-improve-appendix}
    \goodimprove*
        Let $o \in O_g \cap O_{\leq L}$ and let $(o, e(N_o))$ be its local swap from \Cref{thm:2nd-layer}. Let $j \in [L]$ be the index such that $o \in I_j$. Since $o \notin O_s \cup O_d$, there exists an index $i < j$ such that $o \in O^{(i)} \bb (O^{(i)}_d \cup O^{(i)}_s)$. In particular, by Property~\ref{enum:1b} of \Cref{thm:2nd-layer}, we have that $\emptyset \neq e(N_o) \subseteq A_i$. This implies that $m_o = m_{j-1} \leq m_{i}$ so $w_o \leq (1+\gamma)^{-1} m_o \leq (1+\gamma)^{-1} m_i < (1+\gamma)^{-1} c(N_o)$.

    \subsection{Proof of \texorpdfstring{\Cref{thm:APX-easy}}{thm} and \texorpdfstring{\Cref{thm:intro}}{thm}}
    \label{sec:final-approx-value-apdx}
    \APXsimple*
        Recall that $O_{\leq L} \triangleq \lc o \in O \colon w_o \geq m_L \rc$. 
        The proof follows from the fact that the probability of being in $O_b$ is small (\Cref{thm:delta-close}).
        As a consequence,
        \begin{align*}
           \lb 1 - \frac{\gamma}{\e(1+\gamma)} \rb w(O)& \leq \esp{w(O_s \cup O_d \cup O_g)} \\
           & \leq \esp{w(O_d)} + \esp{w(O_g \cap O_{\leq L})} + \esp{w(O_g \bb O_{\leq L})} + \esp{w(A)} \\
           & \leq \esp{w(O_d)} + \esp{w(O_g \cap O_{\leq L})} + \esp{w(O \bb O_{\leq L})} + \esp{w(A)} \\
           & \leq \esp{w(O_d)} + \esp{w(O_g \cap O_{\leq L})} + \delta w(O) + \esp{w(A)},
        \end{align*}
        where the first inequality comes from \Cref{thm:delta-close} and the fact that $O = O_g \sqcup O_b \sqcup O_s \sqcup O_d$.
        The second inequality uses that $w(O_s) \leq w(A)$ ({\Cref{cor: weight Os at most w(A)}}). The last inequality uses \Cref{thm:discarding-OPT}.
        Next, recall that $(1+\gamma) w_o \leq c(N_o)$ for every $o \in O_d\cup (O_g \cap O_{\leq L})$ whenever $(1+\gamma) \leq 2(1-\e)$ by \Cref{thm:2nd-layer} and \Cref{thm:good-improve}. Then by \Cref{thm:discarding-OPT},
        \begin{align*}
            w(O) + \lb \gamma\lb 1 - \frac{\gamma}{\e(1+\gamma)} \rb - \delta \rb  w(O) & \leq w(O) + \gamma \lb \esp{w(O_d)} + \esp{w(O_g \cap O_{\leq L})} + \esp{w(A)}\rb  \\
            & = (1+ \gamma)\lb \esp{w(O_d)} + \esp{w(O_g \cap O_{\leq L})} \rb  \\
            & \quad +\gamma \cdot \esp{w(A)} + \esp{w(O_b \cup O_s \cup (O_g \bb O_{\leq L})} \\
            & \leq \displaystyle\sum_{o \in O} \esp{c(N_o)} + \gamma\esp{w(A)} \\
            & \leq k \cdot \esp{w(A)} + \gamma \esp{w(A)},
        \end{align*}
        where the last inequality uses \Cref{cor: k approximation factor}.
        Rearranging the terms yields the desired result. 

    
    \begin{restatable}{mycor}{APXsimplevalue}
        \label{thm:APX-easy-value}
        Let $A$ be the output of Algorithm~\ref{alg:Random-Threshold}. Then,
        \begin{align*}
            w(O) & \leq \frac{k + 0.2253}{1.118} \cdot \esp{w(A)} 
        \end{align*}
    \end{restatable}
    \begin{proof}[Proof of \Cref{thm:APX-easy-value}]
        Substituting the following values for $\e =0.3873$ and $\gamma =0.2253$ and $\delta = .0001$ we obtain the desired result.
        We check that $\gamma \leq (1 - \e)^{-1} - 1$ and that $ \gamma \leq 2(1-\e) - 1$.
        We have that $(1- \e)^{-1} - 1 > 0.6322$ and $2(1-\e) - 1> 0.2254 $.
        On the other hand, $1+ \gamma \lb 1- \frac{\gamma}{\e(1+\gamma)}\rb - \delta \geq 1.118$. 
    \end{proof}
    Note that as $1/1.118 \leq \frac{9}{10}$ \Cref{thm:APX-easy-value} guarantees the approximation factor \Cref{thm:intro}.

    \section{Improved \texorpdfstring{$(k+1) / (2\ln 2)$}{k} approximation ratio}
\label{sec:Randomized-Threshold}

    In \Cref{sec: remaining edges of O}, we defined the sets $O_b$ and $O_g$ as a function of $\gamma$ measuring the distance between $w_o$ and the closest marker of greater weight $m_o$.
    We have split $O \bb (O_s \cup O_d)$ into the set of $o \in O$ such that $w_o \geq (1+ \gamma^{-1}) m_o$ and those for which $w_o < ( 1+\gamma)^{-1} m_o$. However, an optimal edge is only bad if it is very close to the marker, otherwise a small factor improvement can still be made in the analysis. Given this, we analyze the probability that any $o$ falls within any prescribed range from $m_o$.
    This motivates the definition of the following sets.
    Let $m \in \NN$ and let $\delta_j = \frac{\e j}{m - \e j}$ for $j = 0, \ldots, m$.
    \begin{mydef}
        For $j \in [m]$, we define $R_j \triangleq \lc o \in O \colon w_o \in ((1 + \delta_j)^{-1}, (1+\delta_{j-1})^{-1}] \cdot m_o \rc$ as the set of optimal edges in a factor between $\delta_j$ and $\delta_{j-1}$ close to their closest larger weight marker.
    \end{mydef}
    \begin{remark}
        \label{rmk:T_j-partition-O}
        {Observe that for any $o \in O_{\leq L}$, there exists an index $j \in [m]$ such that $o \in R_j$}. This is by definition of $\delta$ as $\delta_0 = 0$ and $\delta_m = \frac{\e}{1-\e}$. In particular, we get $(1+ \delta_m)^{-1} = 1- \e$ which is equal to  multiplicative gap between two consecutive markers.
    \end{remark}
    Observe that the set $R_j$ is a random set due to the random choice to $\tau$.
    We show that we can derive stronger swap properties than those derived in \Cref{thm:2nd-layer} and \Cref{thm:delta-close}.
    \begin{mylemma}
        \label{thm:improved-bound-double-rest}
        Let $\{(o, e(N_o))\}_{o \in O}$ be the local swaps given by \Cref{thm:2nd-layer}. Then, for any $j \in [m]$ and any $o \in (O_{\leq L} \bb O_s) \cap R_j$, we have
        \begin{align*}
            (1+\delta_{j-1}) w_o & \leq c(N_o).
        \end{align*}
    \end{mylemma}
    \begin{proof}[Proof of \Cref{thm:improved-bound-double-rest}]
        Suppose first that $o \in O^{(i)}_d \cap R_j$ for some $i \in [L]$ and let $(o, e(N_o))$ be its local swap given by \Cref{thm:2nd-layer}.
        We have that $\card{N_o} \geq 2$ for all $o \in O^{(i)}_d$.
        Let $v_1, v_2$ be two distinct vertices, and let $a_1, a_2 \subseteq e(N_o) \subseteq A_i$ be the edges incident to $v_1$ and $v_2$. It is possible that $a_1 = a_2$.
        Since $a_j \in A_i$, we have that $m_{i-1} \geq w_{a_j} > m_{i}$ for $j = 1, 2$.
        Additionally, we have that $o \in O^{(i)}_d$, so $o$ must lie in $I_i$ and $m_o = m_{i-1}$. This means that $w_o \leq (1+\delta_{j-1})^{-1} m_{i-1}$, where we used the definition of $R_j$.
            Therefore,
            \begin{align*}
                w_o & = \frac{w_o}{c(v_1) + c(v_2)} \lb c(v_1) + c(v_2) \rb \leq  \frac{(1+\delta_{j-1})^{-1} m_{i-1}}{2m_{i}} c(N_o)
                = \frac{1}{2(1-\e)(1+\delta_{j-1})} c(N_o),
            \end{align*}
            where the second equality follows from $m_{i}/m_{i-1} = (1-\e)$ for $i \in [L]$. This proves that \Cref{thm:improved-bound-double-rest} holds for $o \in O_d \cap R_j$ using that $(2(1-\e))^{-1} \leq 1$ for $\e \in (0, 1/2)$.\\

        Suppose that $o \in (O_{\leq L} \bb (O_d \cup O_s)) \cap R_j$ and let $(o, e(N_o))$ be its local swap given by \Cref{thm:2nd-layer}.
        Let $\ell \in [L]$ be the index such that $o \in I_\ell$. Since $o \notin O_s \cup O_d$, there exists an index $i < \ell$ such that $o \in O^{(i)} \bb (O^{(i)}_d \cup O^{(i)}_s)$. In particular, by \Cref{thm:2nd-layer}, we have that $\emptyset \neq e(N_o) \subseteq A_i$. This implies that $m_o = m_{\ell-1} \leq m_{i}$ so $w_o \leq (1+\delta_{j-1})^{-1} m_o \leq (1+\delta_{j-1})^{-1} m_i < (1+\delta_{j-1})^{-1} c(N_o)$.
    \end{proof}
    Then by \Cref{thm:delta-close}, we get the following result:
    \begin{mycor}
        \label{thm:proba-landing-close}
        Let $\e > 0$ be the parameter from Algorithm~\ref{alg:Random-Threshold}. For any $o \in O$, we have that
        \begin{align*}
            \prob{w_o \geq (1+\delta_j)^{-1} \cdot m_o} \leq \frac{\delta_j}{\e(1+\delta_j)} = \frac{\frac{\e j}{m - \e j}}{\e \lb 1 + \frac{\e j}{m - \e j} \rb} = \frac{j}{m}.
        \end{align*}
    \end{mycor}

    \subsection{Improved Approximation Ratio}
    Compared to the $9(k+1)/10$ approximation that we have shown so far, we obtain an improved approximation ratio arbitrarily close to $(k+1)/(2\ln(2))$.
    \begin{mythm}
        \label{thm:APX-better}
        For any $\eta > 0$, there exists a choice of parameters $\e \in (0, 1)$ and $\delta \in (0, 1)$ for Algorithm~\ref{alg:Random-Threshold} such that its output $A$ satisfies:
        \begin{align*}
            w(O) & \leq \frac{k+1}{2 \ln(2) - \eta} \cdot \esp{w(A)}.
        \end{align*}
    \end{mythm}
    \begin{proof}[Proof of \Cref{thm:APX-better}]
        For any $\e > 0$, define $\delta \triangleq \frac{\e(1-\e)\eta}{-\ln(1-\e)}$. For each $o \in O$, let $(o, e(N_o))$ be the local swap defined by \Cref{thm:2nd-layer}. Then, either $o \in O_s$ or $o \in (O_{\leq L} \bb O_s) \cap R_j$ for some $j \in [m]$ or $o \in O\bb O_{\leq L}$. In the first and last case, we have that $w_o \leq c(N_o)$. In the second case \Cref{thm:improved-bound-double-rest} shows that $(1+\delta_{j-1}) w_o \leq c(N_o)$. By \Cref{rmk:T_j-partition-O} and any outcome $\tau$, we have that
        \begin{align}
            w(O) & = w(O_s) + \sum_{j = 1}^m w((O_{\leq L} \bb O_s) \cap R_j) + w(O \bb O_{\leq L}) \notag\\
                & \leq \sum_{o \in O} c(N_o) - \sum_{j = 1}^m \delta_{j-1} w((O_{\leq L} \bb O_s) \cap R_j)\notag\\
                & \leq k w(A) - \sum_{j = 1}^m \delta_{j-1} w((O_{\leq L}\bb O_s) \cap R_j), \label{eq:1}
        \end{align}
        where in the last inequality, we used that $\sum_{o \in O} c(N_o) \leq k w(A)$ by \Cref{cor: k approximation factor}. Our proof continues by estimating the rightmost term. Let $\barO \triangleq \{o \in O \colon w_o \geq (1-\e)^{L-1} \cdot W \}$. Since $\delta_j \leq 1$ for $\e \in (0, 1/2)$, we have that:
        \begin{align}
            \sum_{j = 1}^m \delta_{j-1} w((O_{\leq L}\bb O_s) \cap R_j) & \geq \sum_{j = 1}^m \delta_{j-1} w(\barO \cap R_j) - w(O_s)  \notag \\
            & \geq \sum_{j = 1}^m \delta_{j-1} w(\barO \cap R_j) - w(A), \label{eq:2}
        \end{align}
        where we used \Cref{cor: weight Os at most w(A)} in the last inequality.
        Let $p_j = \prob{o \in R_j}$ be the probability that $o \in R_j$. Note that this probability depends only on the outcome $\tau$ that determines the placement of the markers and is therefore equal for every $o \in \bar{O}$.
        Taking the expectation we get that,
        $ \sum_{j = 1}^m \delta_{j-1} \esp{w(\barO \cap R_j)} = w(\barO) \cdot \sum_{j = 1}^m p_j \cdot \delta_{j-1}.$
        We now lower bound the quantity $\sum_{j = 1}^m p_j \cdot \delta_{j-1}$.
        By \Cref{thm:proba-landing-close},
        \begin{align*}
            \sum_{i = j}^m p_{i} = 1 - \prob{w_o \geq ( 1  + \delta_{j-1})^{-1}m_o} \geq 1 - \frac{j-1}{m} & \qquad \mbox{ for all } j \in [m].
        \end{align*}
        We claim that $\displaystyle\sum_{j = 1}^m p_j \cdot \delta_{j-1} \geq \displaystyle\sum_{j = 1}^m \frac{ \delta_{j-1}}{m}$ whenever $(p_1, \hdots, p_m)$ satisfy $\displaystyle\sum_{i = j}^m p_i\geq 1 - \frac{j-1}{m}$ for all $j \in [m]$. Suppose otherwise that there exists some $p' \in [0, 1]^m$ that achieves the minimum value of $\displaystyle\sum_{j = 1}^m p_j \cdot \delta_{j-1}$ and that is not $(1/m, \hdots, 1/m)$. Let $\ell$ be the largest index where $p'_\ell \neq 1/m$. Then $p'_\ell > 1/m$ otherwise $\displaystyle\sum_{i = \ell}^m p'_i < 1 - \frac{\ell-1}{m}$.
        Moreover, there exists an index $q $ such that $p'_q < 1/m$ since otherwise the summed value cannot be less than $\displaystyle\sum_{j = 1}^m \frac{\delta_{j-1}}{m}$.
        Let $q$ be the largest such index and observe that $q < \ell$ as otherwise $\displaystyle\sum_{j = q}^m p'_j < 1 - \frac{q-1}{m}$ contradicting the constraint. Let $p''$ be the vector identical to $p'$ except $p''_q = p'_q + \lambda$ and $p''_\ell = p'_\ell - \lambda$ for some small enough $\lambda >0$. Then since $\delta_\ell \geq \delta_q$, we have that summed values are strictly smaller for $p''$ and that $p''$ still satisfies $\displaystyle\sum_{i = j}^m p''_i\geq 1 - \frac{j-1}{m}$ for all $j \in [m]$. This is a contradiction to the minimality of $p'$.
        This implies that:
        \begin{align}
                \sum_{j = 1}^m p_j \delta_{j-1} & \geq \sum_{j = 1}^m \frac{ \delta_{j-1}}{m} \\
                &= \frac{1}{m}  \sum_{j = 0}^{m-1} \frac{\e j}{ m - \e j} \notag\\
                & \geq \frac{1}{m} \int_0^{m-1} \frac{\e x}{m - \e x} dx \notag\\
                & = \frac{-1}{m} \cdot \ld \frac{m \ln(m - \e x)}{\e} + x \rd_{0}^{m-1} \notag\\
                & = \frac{-1}{m} \lb \frac{m}{\e} \ln\lb \frac{m - \e (m-1)}{m} \rb + m-1 \rb \notag\\
                & = - \frac{\ln(1 - \e + \e/m)}{\e} - 1 + \frac{1}{m} \label{eq:3}
        \end{align}
        Taking the expectation of \Cref{eq:2} and substituting \Cref{eq:3} and \Cref{eq:2} into \Cref{eq:1}, we obtain the following result
        \begin{align}
            w(O) & \leq (k+1) \esp{w(A)} - w(\barO) \cdot \lb - \frac{\ln(1 - \e + \e/m) )}{\e} - 1 + \frac{1}{m} \rb  \label{eq:4}
        \end{align}
        Similar to the proof of \Cref{thm:discarding-OPT}, observe that:
        $w(O \bb \barO) \leq \card{E} (1-\e)^{L-1} W  \leq \frac{\delta}{1-\e} w(O)$. Thus, $w(\barO) \geq (1 - \frac{\delta}{1-\e})w(O)$.
        Substituting the above bound, using that $-\frac{\ln(1-x)}{x} \geq 1$ for all $x \in (0, 1)$, and using the definition of $\delta$, we get that, as $m \rightarrow \infty$,
        \begin{align*}
            w(O) + w(\barO) \cdot \lb - \frac{\ln(1 - \e)}{\e} - 1 \rb & \geq w(O) \lb 1 + \lb 1 - \frac{\delta}{1 - \e}\rb \lb - \frac{\ln(1 - \e)}{\e} - 1 \rb \rb \\
            & \geq w(O) \lb \frac{-\ln(1 - \e)}{\e} - \eta \rb \\
            & \geq w(O) \lb \frac{-\ln(1 - \e) - \eta}{\e} \rb,
        \end{align*}
        Substituting the above bound in \Cref{eq:4}, we get that:
        \begin{align*}
            w(O) & \leq \frac{\e}{- \ln(1-\e) - \eta} (k+1) \esp{w(A)}.
        \end{align*}
        Taking the limit as $\e \rightarrow 1/2$ yields the desired result.
    \end{proof}

    \section{Proof of Matroid Properties}
\label{sec:Analysis-SP}
    In this section, we prove \Cref{thm:Rota-non-basis-2} and \Cref{thm:Laminar-exchanges}. Both theorems are based on the following known result about matroid exchanges:
    \begin{restatable}{mythm}{thmone}[Proposition 6, \cite{Lason:2015:List-Coloring}]
            \label{thm:Rota}
            Let $\cM = (V, \cI)$ be a matroid and $A, B$ be bases of $\cM$. Then, for any partition $A_1, \ldots, A_m$ of $A$, there exists a partition $B_1, \ldots, B_m$ of $B$ such for each $i \in [m]$, we have:
            \begin{align*}
               A_i \cup (B \bb B_i) & \mbox{ is a basis of } \cM.
            \end{align*}
    \end{restatable}
        
    \subsection{Proof of \texorpdfstring{\Cref{thm:Rota-non-basis-2}}{thm}}
        We extend the previous \Cref{thm:Rota} to handle the case where $A$ and $B$ aren't necessarily bases.  
        \GeneralizedRota*
        \begin{proof}[Proof of \Cref{thm:Rota-non-basis-2}]
            Suppose first that $\card{S} = \card{T}$.
            If $S$ is a basis then we can apply \Cref{thm:Rota}.
            Suppose then, that $S$, and $T$ aren't bases of $\cM$.
            Let $\cM' = \cM_{\mid S \cup T}$ be the matroid restricted to $S \cup T$. More precisely, $\cM' = \lc A \subseteq S \cup T \colon A \in \cI \rc$.
            Let $B$ be any basis of $\cM'$. By matroid extension property, there exists $D \subseteq T\bb S$ and $D' \subseteq S\bb T$ such that $S \cup D$ and $T \cup D'$ are basis of $\cM$.
            Let $\cM'' = \lc R \subseteq (S \cup T) \bb (D \cup D') \colon R \cup (D \cup D') \in \cM' \rc$ be the matroid $\cM'$ contracted on $D \cup D'$.
            For $R \subseteq S \cup T$, the rank of $R$ is equal to
            \begin{align*}
               \rank_{\cM''}(R) & = \rank_{\cM'}(R \cup (D \cup D')) - \rank_{\cM'}(D \cup D')
            \end{align*}
            Let $n = \rank(\cM')$, then $n = \card{S \cup D} = \card{T \cup D'}$, and let $m = \rank(\cM'')$. Since $D \cup D' \subseteq T \cup D' \in \cI$, we have that $\rank_{\cM'}{(D \cup D')} = 2d$ where $\card{D} = d$.
            The equation above implies that $m = n - 2d$, and observe that $\rank_{\cM''}(S) = \rank_{\cM'}(S \cup D) - \rank_{\cM'}(D \cup D') = n - 2d = m$. This computation shows that $S \bb D'$ is a base of $\cM''$. An identical computation shows that $T \bb D$ is a base of $\cM''$.

            \paragraph*{Constructing the swaps:} We will apply \Cref{thm:Rota} to $S \bb (D \cup D')$ and $T \bb (D \cup D')$ in $\cM''$. Before doing so, we define an arbitrary bijection $\pi: D' \rightarrow D$ between $D'$ and $D$. This is well defined as $\card{D} = \card{D'}$.
            Using \Cref{thm:Rota} with starting partition $S_i' = S_i \bb D'$ of $S \bb D'$, we find a partition $T_1', \ldots, T_m' \subseteq T \bb D$ such that $\card{S_i' } = \card{T_i'}$ and
            $((T \bb D) \bb T_i') \cup S_i' \in \cI''$.
            We extend these swaps to $\cM'$ and define $T_i = T_i' \cup \pi(S_i \cap  D')$.
            Clearly, we have $\card{T_i} = \card{S_i}$. We prove that $(T \bb T_i) \cup S_i \in \cI$.
            Indeed,
            \begin{align*}
               T \bb T_i \cup S_i & =  \lb (T \bb T_i \cup S_i) \cap (D \cup D') \rb  \cup \lb (T \bb T_i \cup S_i) \bb (D \cup D') \rb.
            \end{align*}
            Observe that the second term $((T \bb T_i \cup S_i) \bb (D \cup D') ) \subseteq ((T \bb D)\bb T_i' ) \cup S_i' \in \cI''$ so is independent on the matroid contracted on $D \cup D'$. Additionally, we have that the first term $( (T \bb T_i \cup S_i) \cap (D \cup D')) \subseteq D \cup D'$.
            Combining both observations implies that: $T \bb T_i \cup S_i \in \cI'$ and so is independent in $\cI$.\\

            It remains to prove the theorem whenever $\card{S} \leq \card{T}$.
            Since $T$ has larger size than $S$, there exists a set $T' \subseteq T \bb S$ such that $S \cup (T \bb T') \in \cI$ and $\card{S \cup (T \bb T')} = \card{T}$.
            In particular, $\card{T'} = \card{S}$. Moreover, both $T'$ and $S$ are independent in the matroid $\cM' = (V, \cI')$ contracted on $T \bb T'$. By the first part of the statement, for any partition $S_1, \ldots, S_m$ of $S$, there is a partition $T'_1, \ldots, T_m'$ such that $\card{S_i} = \card{T_i'}$ and such that $(T' \bb T_i') \cup S_i \in \cI'$.
            Therefore, we have that $ (T \bb T') \cup (T' \bb T_i') \cup S_i = T \bb T_i' \cup S_i \in \cI$. This concludes the proof. 
        \end{proof}

    \subsection{Proof of \texorpdfstring{\Cref{thm:Laminar-exchanges}}{thm}}
        \laminar*
        \begin{proof}[Proof of \Cref{thm:Laminar-exchanges}]
           We contract $\cM$ on $T \bb N_S$ and observe that both $S$ and $N_S$ are independent in the contracted matroid. By \Cref{thm:Rota-non-basis-2}, we have that there exists a partition $\{N_{S_i}\}_{i \in [m]}$ such that  $\card{N_{S_i}} = \card{S_i}$ and $N_S \bb N_{S_i} \cup S_i $ is independent in the contracted matroid. In particular, we have that $T \bb N_S \cup \lb N_S \bb N_{S_i} \cup S_i \rb = T \bb N_{S_i} \cup S_i \in \cI$.
        \end{proof}

    \section{Runtime and probability guarantee of Algorithm~\ref{alg:Random-Threshold}}
\label{sec:algorithm-appendix}
    \subsection{Polynomial runtime}
    \label{sec:polytime}
    
    Let $n= \card{E}$. $\SLS$ calls $\ILS$ $L$ times.
    Hence, $\SLS$ runs in $O(L \cdot \card{LS})$-time, where $\card{LS}$ is the run time of $\ILS$. The latter algorithm runs in polynomial time. Finding an $(A, I, I)$-swap uses $O(n^{O(k)})$ calls to the independence matroid oracle. Indeed, for all possible subsets of size $2$, i.e., $O(n^2)$ of them, we verify the existence of a set of size $2k$ that we can discard by checking  $O(n^{2k})$ possible sets.
    Although the number of improving $(A, I, I)$-swaps can be exponential, standard weight-scaling techniques \cite{Lee:2010:Submodular, Lee:2009:Matroid, Berman:2000:d/2} can be used to reduce the number of iterations to ensure polynomial-time convergence and termination. 
    
    
    \subsubsection{Weight rescaling} Suppose that $A$ is a locally-optimal solution with respect to the improvement considered in Algorithm~\ref{alg:local-search}. Let $W \triangleq \max_{e \in E, v(e) \in \cI} w_e$ be the maximum weight in our instance, let $\e > 0$ be an arbitrarily small constant. We apply Algorithm~\ref{alg:Random-Threshold} on the instance where we rescale the weights of the instance as follows. For each $e \in E$, we define $\tilde{w}(e) = \left\lfloor M w(e) \right\rfloor$, where $M \triangleq \frac{n}{\e W}$. \\

    First, we prove that the weight-scaling doesn't affect the approximation guarantee of Algorithm~\ref{alg:Random-Threshold}.
    Let $A$ be the solution output by Algorithm~\ref{alg:Random-Threshold}, let $O$ be the optimal solution and let $\alpha > 0$ be be the approximation factor, such that $\alpha w(A) \geq w(O)$. 
    By definition of $\tilde{w}$, we have that $M w(a) \geq \tilde{w}(a) \geq M w(a) - 1$. 
    Combining the above properties, we get that
    \begin{align*}
        \alpha M w(A) & \geq \alpha \tilde{w}(A) \geq \tilde{w}(O) \geq M w(O) - \card{O} \geq M w(O) - n.
    \end{align*}
    Dividing through by $M$, using that $M = \frac{n}{\e W}$ and that $W \leq w(O)$, we get that
    \begin{align*}
        \alpha w(A) &  \geq w(O) - \frac{n}{M} \geq w(O) - \e W \geq ( 1- \e) w(O).
    \end{align*}
    This proves that the weight-scaling doesn't affect the approximation guarantee of Algorithm~\ref{alg:Random-Threshold}. It remains to prove that the technique makes the local-search run in polynomial-time.

    Here, we use the fact that $\tilde{w}(e)$ is an integer. Indeed, the optimal solution has value at most $\tilde{w}(O) \leq \card{O} M \cdot W = \frac{n^2}{\e}$. Thus, there is at most $n^2 \e^{-1}$ improving $(A, I, I)$-swaps.

    \subsection{Last interval has negligible weight -- Proof of \texorpdfstring{\Cref{thm:discarding-OPT}}{thm}}
    \label{sec:Last-Interval}
    \discardOPT*
        Using that $L = \lceil -\log_{1-\e}\lb \card{E} \delta^{-1} \rb \rceil + 1$, we show that $w(O \bb O_{\leq L})$ is small.
        \begin{align*}
            w(O \bb O_{\leq L}) & \leq |O \bb O_{\leq L}| \cdot m_{L} 
             \leq \card{E}(1 - \e)^{-\log_{1 - \e}(\card{E}\delta^{-1})} (1-\tau) W 
             \leq \delta W 
            \leq \delta w(O),
        \end{align*}
        where we used that $w(O) \geq W$ since $W$ is the largest weight of an independent edge and $(1 - \tau) \leq 1$ for all outcomes $\tau$. 

\subsection{Probability of Algorithm~\ref{alg:Random-Threshold} succeeding}
    \label{sec:high-proba-guarantee}
        We prove that $\SLS$ succeeds with high-probability. Let $A$ be the solution returned by $\SLSS$, and let $\alpha$ be the approximation factor such that $\esp{A} \geq (\alpha/k)\cdot  w(O)$.
        We show that $w(A) \geq (\alpha/k)\cdot  w(O)$ with high probability. 
        This follows from a reverse Markov inequality.
        \begin{mythm}
            \label{thm:reverse-Markov}
            Let $X$ be a random variable such that $\prob{X \leq a} = 1$ for some constant $a$. Then, for any $d < \esp{X}$, the following holds:
            \begin{align*}
                \prob{X > d } & \geq \frac{\esp{X} - d}{a - d}.
            \end{align*}
        \end{mythm}
        Applying \Cref{thm:reverse-Markov} with $a = w(O)$ and $d = \alpha (1- \xi)/k w(O)$ for some small parameter $\xi$, we get that
        \begin{align*}
            \prob{X > \alpha (1- \xi)/k \cdot w(O)} & \geq \frac{\alpha \xi}{k - \alpha(1- \xi)}.
        \end{align*}
        Running $O(k/\xi)$ independent copies of $\SLS$ and taking the best solution ensures that the expected guarantee of $\SLS$ holds with probability arbitrarily close to $1$.

    \section{Reduction and Simplifying Assumptions}
\label{sec:Reduction}
\subsection{Reduction from \texorpdfstring{$k$}{k}-Matroid Intersection to Matroid \texorpdfstring{$k$}{k}-Parity}
For completeness, we provide the reduction from $k$-Matroid Intersection to Matroid $k$-Parity as shown by Lee, Sviridenko and Vondr\'ak \cite{Lee:2009:Matroid}.
Let $\Pi$ be an instance of $k$-Matroid Intersection. We are given $k$-matroid $\cM_1 = (V, \cI_1), \ldots, \cM_k = (V, \cI_k)$ on the same groundset $V$. We would like to construct an instance $\Pi' = (G = (V', E), \cM)$ of Matroid $k$-Parity.  \\

The set of vertices $V'$ consists of $k$ copies of $V$, which we denote by $V_1, \ldots, V_k$. For each copy, we define the matroid $\cM'_i = (V_i, \cI_i)$ that is the matroid $\cM_i$ on the $\nth{i}$ copy of $V$.
Our hypergraph $G = (V', E)$ then contains $n = \card{V}$ parallel (disjoint) hyperedges on the $k$ copies of the same element from $V$.
Observe that a collection of hyperedges such that, for all $i \in [k]$, the set of incident vertices in the $\nth{i}$ copy is independent defines an independent set in the intersection of $k$ matroids and vice-versa. \\

It remains to prove that this property can be encoded as a single matroid. In fact, we simply define $\cM$ as the \emph{union} of the matroid $\cM'_1, \ldots, \cM'_k$. More precisely, we let $\cM = \cM'_1 \vee \ldots \vee \cM'_k$ defined on the groundset $V'$ so that
$\cI = \{ I_1 \cup \ldots \cup I_k \colon I_j \in \cI_j \mbox{ for all } j \in [k] \}$. In contrast to matroid intersection, $\cM$ is in fact a matroid (See \cite{Schrijver:2003:Combinatorial} for a proof).

\subsection{Disjointness Assumption}
\label{sec:Disjointness}
    In this section, we show that, given an instance of $k$-Matroid Parity $\Pi = (G = (V, E), \cM)$, we may assume that each vertex belongs to a unique hyperedge. This assumption is proved in \cite{Lee:2009:Matroid}. We recall the proof here.
    Let $\Pi$ be our $k$-Matroid Parity instance. We create a novel instance $\Pi' = (G' = (V', E'), \cM')$ equivalent to $\Pi$.
    For each vertex $v \in V$, create $n_v \in V'$ copies of $v$, where $n_v$ is the degree of $v$ in the hypergraph $G$. We replace each hyperedge in $G$ by a collection of distinct copies of its elements, so that the new hyperedges are disjoint.
    It remains to define $\cM'$. A set $S' \subseteq V'$ is independent in $\cM'$ if $S$ contains at most $1$ copy of each vertex $V$ and the respective set $S \subseteq V$ is independent in $\cM$.
    It is simple to check that $\cM'$ is a matroid.

\end{document}